\documentclass[12pt,twoside,a4paper]{article}
\usepackage{amsmath,amssymb,latexsym,theorem,natbib,epsfig,color,subfigure,footmisc,bbm}
\usepackage{multirow,graphicx,array,rotating,url}  
\usepackage{epstopdf,afterpage,wasysym}
\usepackage{verbatim}
\usepackage[normalem]{ulem}
\def\crps{\mathop{\hbox{\rm CRPS}}}

\def\bs{\mathop{\hbox{\rm BS}}}

\numberwithin{equation}{section}

\title{Parametric model for post-processing visibility ensemble forecasts}

\author{{\'Agnes Baran} and {S\'andor Baran}$^{*}$  \vspace*{0.5cm}\\
{\small Faculty of Informatics, University of Debrecen, Hungary}
}  

\date{}

\begin{document}

\maketitle

\footnotetext[1]{Corresponding author: \url{baran.sandor@inf.unideb.hu}}
\begin{abstract}
Although by now the ensemble-based probabilistic forecasting is the most advanced approach to weather prediction, ensemble forecasts still might suffer from lack of calibration and/or display systematic bias, thus require some post-processing to improve their forecast skill. Here we focus on visibility,  which quantity plays a crucial role e.g. in aviation and road safety or in ship navigation, and propose a parametric model where the predictive distribution is a mixture of a gamma and a truncated normal distribution, both right censored at the maximal reported visibility value. The new model is evaluated in two case studies based on visibility ensemble forecasts of the European Centre for Medium-Range Weather Forecasts covering two distinct domains in Central and Western Europe and two different time periods. The results of the case studies indicate that post-processed forecasts are substantially superior to the raw ensemble; moreover, the proposed mixture model consistently outperforms the Bayesian model averaging approach used as reference post-processing technique.

\bigskip
\noindent {\em Keywords:\/} Bayesian model averaging, ensemble calibration, predictive distribution, visibility
\end{abstract}

\section{Introduction}
\label{sec1}

Despite the continuous improvement of autoland, autopilot, navigation and radar systems, visibility conditions are still critical in  aviation and road safety and in ship navigation as well. Nowadays, visibility observations are obtained automatically; visibility sensors take the measurements of ``the length of atmosphere over which a beam of light travels before its luminous flux is reduced to 5\,\% of its original value\footnotemark[2]'', which quantity is called meteorological optical range. 
\footnotetext[2]{\url{https://www.metoffice.gov.uk/weather/guides/observations/how-we-measure-visibility} [Accessed on 22 January 2024]}

Visibility forecasts are generated with the help of numerical weather prediction (NWP) models either as direct model outputs or by utilizing various algorithms \citep[see e.g.][]{sw99,gmb06,wkl23} to calculate visibility from forecasts of related weather quantities such as, for instance, precipitation or relative humidity \citep{cr11}. Nowadays, the state-of-the-art approach to weather prediction is to issue ensemble forecasts by running an NWP model several times with perturbed initial conditions or different model parametrizations \citep{btb15,b18a}. Hence, for a given location, time point and forecast horizon, instead of having a point forecast, a forecast ensemble is issued. It opens the door for estimating the forecast uncertainty  or even the probability distribution of the future weather variable \citep{gr05}, and provides an important tool for forecast-based decision making \citep{ffh19}. In particular, several recent studies \citep[see][]{pmtar21,pgd22} verify the superiority of probabilistic predictions e.g. in fog forecasting, which is one of the most frequent reasons of low visibility. 

By now, all major weather centres operate ensemble prediction systems (EPSs); however, only a few has visibility among the forecasted parameters. For instance, since 2015, visibility is part of the Integrated Forecast System \citep[IFS;][]{ifs21} of the European Centre for Medium-Range Weather Forecasts \citep[ECMWF;][]{ecmwf12}; nevertheless, it is an experimental product and ``expectations regarding the quality of this product should remain low'' \citep[][Section 9.4]{oh18}. A further example is the Short-Range Ensemble Forecast System of the National Centers for Environmental Prediction, which covers the Continental US, Alaska, and Hawaii regions \citep{zdmqd09}.

A typical problem with the ensemble forecasts is their underdispersive and biased feature, which has been observed with several operational EPSs \citep[see e.g.][]{bhtp05} and can be corrected with some form of post-processing \citep{b18b}. In the last decades a multitude of statistical calibration techniques have been proposed for a broad range of weather parameters; see \citet{w18} or \citet{vbd21} for an overview of the most advanced approaches. Non-parametric methods usually represent predictive distributions via their quantiles estimated by some form of quantile regression \citep[see e.g.][]{fh07,brem19}, whereas parametric models such as Bayesian model averaging \citep[BMA;][]{rgbp05} or ensemble model output statistics \citep[EMOS;][]{grwg05} provide full predictive distributions of the weather variables at hand. The BMA predictive probability density function (PDF) of a future weather quantity is the weighted sum of individual PDFs corresponding to the ensemble members, which form of the predictive PDF might be beneficial in situations when multimodal predictive distributions are required \citep[see e.g.][]{bhea19}. In contrast, the EMOS (also referred to as non-homogeneous regression) predictive distribution is given by a single parametric family, where distributonal parameters are given functions of the ensemble members.   Furthermore, recently machine learning-based approaches gain more and more popularity in ensemble post-processing both in parametric framework \citep[see e.g.][]{rl18,gzshf21,bb24} and in non-parametric context \citep{brem20}; for a systematic overview of the state-of-the-art techniques see \citet{sl22}. Finally, in the case of discrete quantities, such as total cloud cover (TCC), the predictive distribution is a probability mass function and post-processing can be considered as a classification problem, where both parametric techniques \citep{hhp16} and advanced machine learning-based classifiers can be applied \citep{bleab21}.

Although, as mentioned, visibility forecasts are far less reliable than ensemble forecasts of other weather parameters \citep[see e.g.][]{zdgd12}, only a few of the above mentioned methods is adapted to this particular variable. \citet{cr11} consider a BMA approach where each individual predictive PDF consists of a point mass at the maximal reported visibility and a beta distribution, which models the remaining visibility values. \citet{rh18} propose a non-parametric method for calibrating short-range visibility predictions obtained using the Weather Research and Forecasting Model \citep{rh14}. Furthermore, since most synoptic observation (SYNOP) stations report visibility in discrete values according to the WMO suggestions, in a recent study \citet{bl23} investigate the  approach of \citet{hhp16} and \citet{bleab21}, and obtain (discrete) predictive distributions of visibility with the help of proportional odds logistic regression and multilayer perceptron neural network classifiers.

In the present article, we develop a novel parametric post-processing model for visibility ensemble forecasts where the predictive distribution is a mixture of a gamma and a truncated normal distribution, both right censored at the maximal reported visibility value. The proposed mixture model is applied in two case studies that focus on ECMWF visibility ensemble forecasts covering two distinct
domains in Central and Western Europe and two different time periods. As reference post-processing approach we consider the BMA model of \citet{cr11}; nonetheless, we report the predictive performance of climatological and raw ensemble forecasts as well. 

\begin{figure}[t]
   \centering
   \epsfig{file=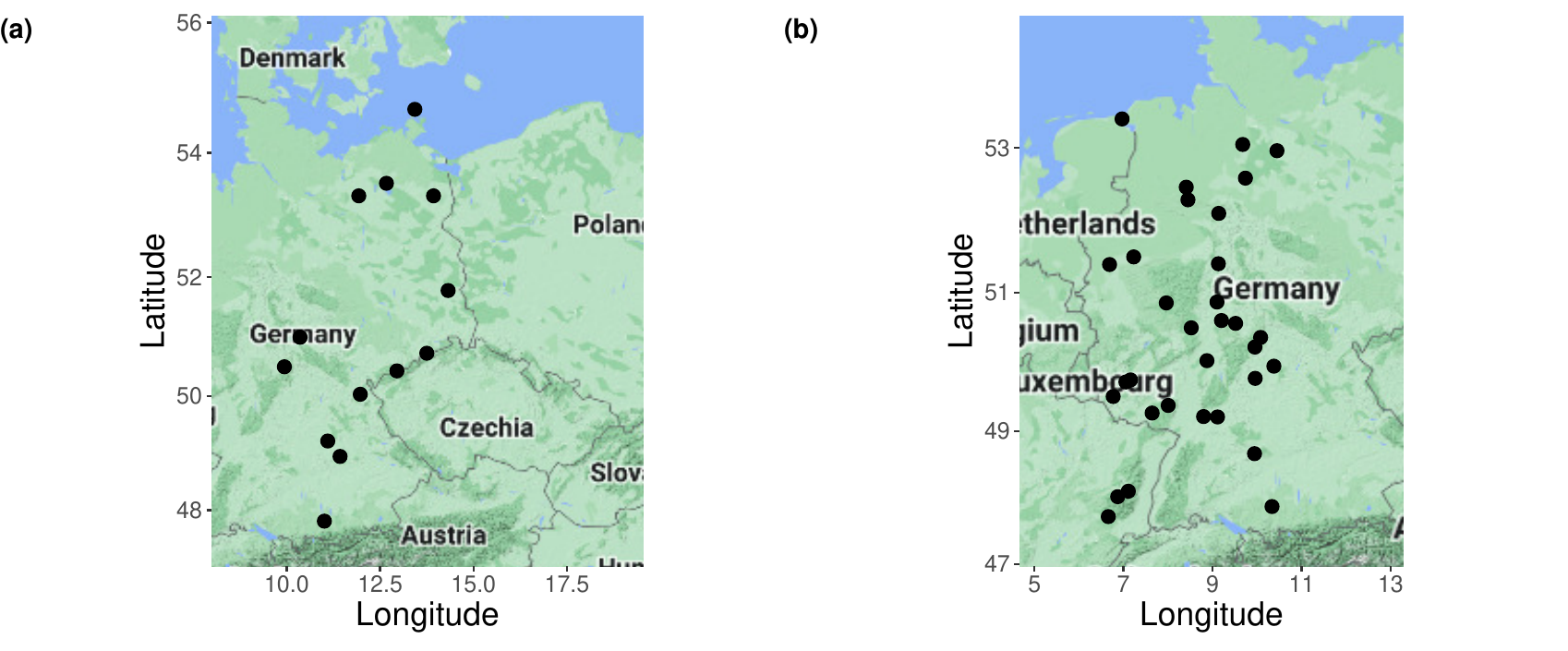, width=\textwidth} 
   \caption{Locations of SYNOP observation stations  corresponding to (a) ECMWF forecasts for 2020-2021; (b) EUPPBench benchmark dataset.}
   \label{fig:map}
 \end{figure}

The paper is organized as follows. Section \ref{sec2} briefly introduces the visibility datasets considered in the case studies. The proposed mixture model, the reference BMA approach, training data selection procedures and tools of forecast verification are provided in Section \ref{sec3}, followed by the results for the two case studies presented in Section \ref{sec4}. Finally, concluding remarks and lessons learned can be found in Section \ref{sec5}.

\section{Data}
\label{sec2}

\begin{table}[t]
  \begin{center}
    \begin{tabular}{l|c|c}
      &ECMWF data for 2020--2021& {\em EUPPBench\/} benchmark data\\ \hline
      &&high resolution forecast (HRES) \\ \cline{2-3}
Ensemble members  &\multicolumn{2}{c}{control forecast (CTRL)}\\\cline{2-3}
      &\multicolumn{2}{c}{50 members (ENS) generated using perturbations}\\ \hline
      Period (calendar years)&2020--2021&2017--2018\\ \hline
      No. of SYNOP stations&13&32\\ \hline
      Forecast horizon&240 h&120 h\\ \hline
      Times step&\multicolumn{2}{c}{6 h} \\ \hline
      Initialization&\multicolumn{2}{c}{0000 UTC} \\ \hline
      Missing observations&none&around 2\,\% \\ \hline
      Missing forecasts&2 forecast cases& none       
 \end{tabular}
\end{center}
\caption{Overview of the studied datasets}
\label{tab:datasets}
\end{table}

 In the case studies of Section \ref{sec4} we evaluate the mixture model proposed in Section \ref{subs3.1} using ECMWF visibility ensemble forecasts (given in 1 m steps) and corresponding validating observations (reported in 10 m increments) covering two different time periods and having disjoint but geographically close ensemble domains. In fact, we consider subsets of the datasets studied in \citet{bl23} by selecting only those locations where the resolution of the reported observations is close to that of the forecasts and can be considered as continuous. The first dataset comprises the operational 51-member ECMWF visibility ensemble forecasts for calendar years 2020 and 2021, whereas the second contains  visibility data of the {\em EUPPBench\/} benchmark dataset \citep{eupp} for calendar years 2017 -- 2018. The locations of the investigated SYNOP stations are given in Figure \ref{fig:map}, while
Table \ref{tab:datasets} provides an overview of both studied datasets.

\section{Parametric post-processing of visibility}
\label{sec3}
As mentioned in the Introduction, EMOS is a simple and efficient tool for post-processing ensemble weather forecasts \citep[see also][]{vbd21}. However, as it fits a single probability law to the forecast ensemble chosen from a given parametric distribution family, EMOS is usually not flexible enough to model multimodal predictive distributions. A natural approach is to consider a mixture of several probability laws, which is also the fundamental idea of the BMA models.
Furthermore, visibility is non-negative and the reported observations are often limited to a certain value (in our case studies to 75 and 70 km), which restriction should be taken into account, too. A possible solution is to censor a non-negative predictive distribution from above or to mix a continuous law and a point mass at the maximal reported visibility value. The former approach appears in the mixture model proposed in Section \ref{subs3.1}, whereas the reference BMA model of \citet{cr11} described briefly in Section \ref{subs3.2} is an example of the latter.

In the following sections, let \ $f_1,f_2,\ldots ,f_{52}$ \ denote a 52-member ECMWF visibility ensemble forecasts for a given location, time point and forecast horizon, where \ $f_1=f_{HRES}$ \ and  \ $f_2=f_{CTRL}$ \ are the high-resolution and control members, respectively, whereas \ $f_3,f_4, \ldots ,f_{52}$ \ correspond to the 50 members generated using perturbed initial conditions. These members, which we will denote by \ $f_{ENS,1},f_{ENS,2}, \ldots, f_{ENS,50}$, \  lack individually distinguishable physical features, hence they are statistically indistinguishable and can be treated as exchangeable. In what follows, \ $\overline f_{ENS}$ \ and \ $S_{ENS}$ \ will denote the mean and standard deviation of the 50 exchangeable ensemble members, respectively, and following the suggestions of e.g. \citet{frg10} or \citet{w18}, in the models presented in Sections \ref{subs3.1} and \ref{subs3.2} these members will share the same parameters.

\subsection{Mixture model}
\label{subs3.1}

\begin{figure}[t]
   \centering
   \epsfig{file=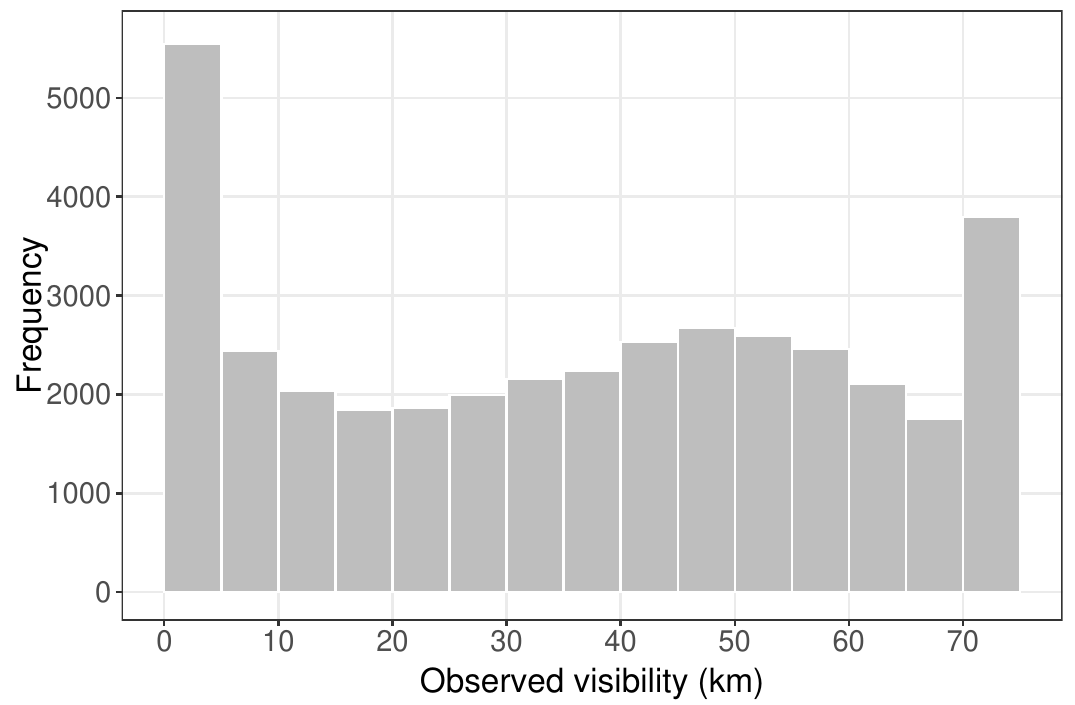, width=.6\textwidth} 
   \caption{Climatological frequency histogram of visibility for calendar years 2020 -- 2021.}
   \label{fig:obs}
 \end{figure}

 According to the climatological histogram of Figure \ref{fig:obs}, a unimodal distribution is clearly not appropriate to model visibility. One has to handle separately low visibility values, there is a second hump at medium to large visibility and a censoring is required at the maximal reported value \ $x_{\max}$. \

 Let \ $g(x\vert \kappa,\theta)$ \ and \  $G(x\vert \kappa,\theta)$ \ denote the probability density function (PDF) and cumulative distribution function (CDF) of a gamma distribution \ $\Gamma (\kappa,\theta)$ \ with shape \ $\kappa >0$ \ and scale \ $\theta>0$, \ respectively, while notations \ $h(x\vert \mu,\sigma^2)$ \ and \ $H(x\vert \mu,\sigma^2)$ \ are used for the PDF and the CDF of a normal distribution \ $\mathcal N_0(\mu,\sigma^2)$ \ with location \ $\mu$, \ scale \ $\sigma>0$ \ left truncated at zero. Furthermore, denote by  \ $g^c(x\vert \kappa,\theta)$ \ and  \ $h^c(x\vert \mu,\sigma^2)$ \ the PDFs of the censored versions of these laws, that is
 \begin{align*}
   g^c(x\vert \kappa,\theta)&:=g(x\vert \kappa,\theta) {\mathbb I}_{\{x<x_{\max}\}} + \big(1-G(x_{\max}\vert \kappa,\theta)\big) {\mathbb I}_{\{x= x_{\max}\}}, \\
    h^c(x\vert \mu,\sigma^2)&:=h(x\vert \mu,\sigma^2) {\mathbb I}_{\{x<x_{\max}\}} + \big(1-H(x_{\max}\vert \mu,\sigma^2)\big) {\mathbb I}_{\{x=x_{\max}\}},
 \end{align*}
where \ $\mathbb I_A$ \ denotes the indicator function of a set \ $A$.

The proposed predictive distribution of visibility is a mixture of censored gamma and censored truncated normal distributions
\begin{equation}
  \label{eq:mixPDF}
  p(x|\kappa,\theta,\mu,\sigma^2,\omega)=(1-\omega)g^c(x\vert \kappa,\theta) + \omega  h^c(x\vert \mu,\sigma^2),
  \end{equation}
  where both the weight \ $\omega\in [0,1]$ \ and the parameters of the component distributions depend on the ensemble forecast. In particular,
  \begin{equation*}
    \omega = 1/\big(1+\exp(-\gamma \overline f_{ENS})\big),
  \end{equation*}
which is a smooth monotone function of \ $\overline f_{ENS}$. \ Furthermore,
the mean \ $m=\kappa\theta$ \ and variance \ $v=\kappa\theta^2$ \ of the uncensored gamma distribution \ $\Gamma (\kappa,\theta)$ \ are given as
\begin{equation*}
  m=a_0+a_1^2f_{HRES}+a_2^2f_{CTRL} +a_3^2\overline f_{ENS} + a_{4}B_1(d) +  a_{5}B_2(d) \quad \text{and} \quad v=b_0 + b_1^2 S^2_{ENS},
\end{equation*}
while location and scale of the truncated normal distribution \  $\mathcal N_0(\mu,\sigma^2)$ \ are expressed as
\begin{equation*}
  \mu=\alpha_0+\alpha_1^2f_{HRES} +\alpha_2^2f_{CTRL} +\alpha_3^2\overline f_{ENS} + \alpha_{4}B_1(d) +  \alpha_{5}B_2(d) \quad \text{and} \quad \sigma=\beta_0 + \beta_1^2 S_{ENS},
\end{equation*}
where functions \ $B_1(d)$ \  and \ $B_2(d)$ \ are annual base functions
\begin{equation}
  \label{eq:abf}
     B_1(d):=\sin\big(2\pi d/365) \qquad \text{and} \qquad B_2(d):=\cos\big(2\pi d/365)
   \end{equation}
addressing seasonal variations in the mean/location \citep[see e.g.][]{dmmz17} and \ $d$ \ denotes the day of the year. Following the optimum score principle of \citet{gr07}, model parameters \ $\gamma, a_0,a_1, \ldots ,a_{5}, \alpha_0, \alpha_1, \ldots, \alpha_{5},b_0,b_1,\beta_0,\beta_1 \in{\mathbb R}$ \ are estimated by optimizing the mean value of an appropriate proper scoring rule, namely the logarithmic score (see Section \ref{subs3.4}), over an appropriate training data set comprising past forecast-observation pairs. In the case when the high-resolution and/or the control forecast is not available, one obviously sets \ $a_1=\alpha_1=0$ \ and/or \ $a_2=\alpha_2=0$.
    
\subsection{Bayesian model averaging}
\label{subs3.2}
The BMA predictive distribution of visibility \ $X$ \  based on 52-member ECMWF ensemble forecasts is  given by
\begin{equation}
  \label{eq:bma}
  \mathfrak p\big(x|f_1,f_2,\ldots ,f_{52}, \boldsymbol\theta_1,\boldsymbol\theta_2,\ldots, \boldsymbol\theta_{52}\big)=\sum_{k=1}^{52}\omega_k\mathfrak h\big(x|f_k;\boldsymbol\theta_k\big),
\end{equation}
where \ $\omega_k$ \ is the weight and \ $\mathfrak h\big(x|f_k;\boldsymbol\theta_k\big)$ \ is the component PDF corresponding to the $k$th ensemble member with parameter vector \ $\boldsymbol\theta_k$ \ to be estimated with the help of the training data. Note that weights form a probability distribution  \ ($\omega_k\geq 0, \ k=1,2, \ldots ,52$, \ and \ $\sum_{k=1}^{52}\omega_k=1$) \ and \ $\omega_k$ \ represents the relative performance of the forecast \ $f_k$ \ in the training data. 

In the BMA model of \citet{cr11} the conditional PDF  \ $\mathfrak h\big(x|f_k;\boldsymbol\theta_k\big)$ \ is based on the square root of the forecast \ $f_k$ \ and consists of two parts. The first models the point mass at the maximal reported visibility \ $x_{\max}$ \ using logistic regression as
\begin{equation}
  \label{eq:logit}
{\rm logit}\,{\mathsf P}\big(X=x_{\max}|f_k\big) = \log \frac{{\mathsf P}\big(X=x_{\max}|f_k\big)}{{\mathsf P}\big(X<x_{\max}|f_k\big)} = \pi_{0k} + \pi_{1k}f_k^{1/2}.
  \end{equation}

The second part provides a continuous model of visibility given that it is less than \ $x_{max}$ \ using a beta distribution with shape parameters \ $\alpha,\beta >0$ \ and support \ $[0,x_{\max}]$ \ defined by PDF 
\begin{equation*}
  q(x|\alpha,\beta):=\frac {\big(y/x_{\max}\big)^{\alpha-1} \big(1-y/x_{\max}\big)^{\beta-1}}{{\mathcal B}(\alpha,\beta)x_{\max}}, \qquad x \in [0,x_{\max}], 
\end{equation*}
where \ ${\mathcal B}(\alpha,\beta)$ \ is the beta function. Given the $k$th ensemble member \ $f_k$, \ the mean \ $x_{\max}\,\alpha /(\alpha + \beta)$ \ and standard deviation \ $ x_{\max}\sqrt{\alpha\beta} /\big((\alpha+\beta) \sqrt{\alpha+\beta+1}\big)$ \ of the corresponding beta distribution are expressed as
\begin{equation}
  \label{eq:betaPars}
  \mathfrak m_k=\varrho_{0k}+\varrho_{1k} f_k^{1/2} \qquad \text{and} \qquad
  \mathfrak s_k=c_{0}+c_{1}f_k^{1/2},
\end{equation}
respectively. Note that variance parameters in \eqref{eq:betaPars} are kept constant for practical reasons. On the one hand, this form reduces the number of unknown parameters to be estimated and helps in  avoiding overfitting. On the other hand, as argued by \citet{srgf07} and \citet{cr11}, in a more general model allowing member-dependent variance parameters \ $c_{0k}$ \ and \ $c_{1k}$, \ these parameters do not vary much from one forecast to another.

Now, the conditional PDF of visibility given the $k$th ensemble member \ $f_k$ \ is
\begin{equation*}
 \mathfrak h\big(x|f_k;\boldsymbol\theta_k\big)= {\mathsf P}\big(X<x_{\max} | f_k\big)\mathfrak q(x|f_k){\mathbb I}_{\{X<x_{\max}\}} + {\mathsf P}\big(X=x_{\max} | f_k\big){\mathbb I}_{\{X=x_{\max}\}}
  \end{equation*}
  where \ ${\mathsf P}\big(X=x_{\max} | f_k\big)$ \ is defined by \eqref{eq:logit}, \ $\mathfrak q(x|f_k)$ \ denotes the beta distribution with mean and standard deviation specified by \eqref{eq:betaPars} and \ $\boldsymbol\theta_k:=\big(\pi_{0k},\pi_{1k},\varrho_{0k},\varrho_{1k},c_0, c_1\big)$. 

  Parameters \ $\pi_{0k}$ \ and \ $\pi_{1k}$ \ are estimated from the training data by logistic regression, mean parameters \ $\varrho_{0k}$ \ and $\varrho_{1k}$ \  are obtained using linear regression connecting the visibility observations less than \ $x_{\max}$ \ to the square roots of the corresponding ensemble members, whereas for estimating weights \ $\omega_k$ \ and variance parameters \ $c_0, \ c_1$ \ one uses the maximum likelihood approach with EM algorithm to maximize the likelihood function. For more details we refer to \citet{cr11} and note that following again \citet{frg10}, we do not distinguish between the exchangeable ensemble members \ $f_3,f_4, \ldots ,f_{52}$ \ and assume \ $\omega_3=\omega_4=\cdots =\omega_{52}$ \ and \ $\boldsymbol\theta_3=\boldsymbol\theta_4=\cdots =\boldsymbol\theta_{52}$. \ Furthermore, if some of the ensemble forecasts are missing, then the corresponding weights should be set to zero.

\subsection{Temporal and spatial aspects of training}
\label{subs3.3}

The parameters of the mixture and BMA predictive PDFs described in Sections \ref{subs3.1} and \ref{subs3.2}, respectively, are estimated separately for each individual lead time. For a given day \ $d$ \ and lead time \ $\ell$ \ the estimation is based on training data (observations and matching forecasts with the given lead time) from an $n$-day long time interval betweeen calendar days \ $d-\ell -n+1$ \ and  \ $d-\ell$, \ that is one considers data of the latest \ $n$ \ calendar days when the date of validity of the $\ell$-day ahead forecasts preceds the actual day  \ $d$. \ The optimal length of the rolling training period is determined by comparing the predictive performance of post-processed forecasts for various lengths \ $n$.

As in both investigated datasets consist of forecast-observation pairs for several SYNOP stations, one can consider different possibilities for the spatial composition of the training data. The simplest and most parsimonious approach is the regional modelling \citep{tg10}, where all investigated locations are treated together and they share a single set of model parameters. Regional models allow an extrapolation of the predictive distribution to locations where only forecasts are available \citep[see e.g.][]{bb24}; nonetheless, if the ensemble domain is too large and the stations have quite different characteristics, then this approach is not really suitable and, as demonstarted e.g. by \citet{lb17} or \citet{bbpbb20}, might even fail to outperform the raw ensemble. In contrast, local models result in distinct parameter estimates for different locations as they are based only on the training data of each particular site.  In this way, one can capture local characteristics better, so local models usually outperform their regional counterparts, as long as the amount of training data is large enough. Thus, one needs much longer training windows than in the regional case. For instance, \citet{hspbh14} suggest 720-, 365- and 1816-day rolling training periods for EMOS modelling of temperature, wind speed and precipitation accumulation, respectively. Finally, the advantages of regional and local parameter estimation can be combined by the use of semi-local techniques, where either training data of a given location is augmented with data of sites with similar characteristics, or the ensemble domain is divided into more homogeneous subdomains and, within each subdomain, a regional modelling is performed. In the case studies of Section \ref{sec4}, besides local and regional parameter estimation, we also consider the clustering-based semi-local approach suggested by \citet{lb17}. For a given date of the verification period, to each observation station we first assign a feature vector depending both on the station climatology and the forecast errors of the raw ensemble mean over the training period. Then, based on the corresponding feature vectors, the stations are grouped into homogeneous clusters using $k$-means clustering and dynamically regrouped as the training period rolls ahead.

\subsection{Model verification}
\label{subs3.4}

Forecast skill is advised to be evaluated with the help of proper scoring rules \citep[see e.g.][]{gr07}, which can be considered as loss functions aiming to maximize the concentration (sharpness) of the probabilistic forecasts subject to their statistical consistency with the corresponding observations (calibration). One of the most used proper scoring rules is
the logarithmic score \citep[LogS;][]{good52}, that is the negative logarithm of the predictive PDF evaluated at the validating observation. The other very popular proper score is the continuous ranked probability score \citep[CRPS;][Section 9.5.1]{w19}. For a forecast provided in the form of a CDF \ $F$ \ and a real value \ $x$ \ representing the verifying observation, the CRPS is defined as
\begin{equation}
    \label{eq:CRPSdef}
\crps(F,x) := \int_{-\infty}^{\infty}\big[F(y)-{\mathbb I}_{\{y\geq x\}}\big]^2{\mathrm d}y ={\mathsf E}|X-x|-\frac 12
{\mathsf E}|X-X'|,
\end{equation}
where \ ${\mathbb I}_H$ \ denotes the indicator function of a set \ $H$, while \ $X$ \ and \ $X'$ \ are independent random variables distributed according to \ $F$ \ with finite first moment. Both the LogS and the CRPS are negatively oriented scores, that is smaller values mean better predictive performance. In most applications the CRPS has a simple closed form \citep[see e.g.][]{jkl19}; however, this is not the case for the predictive distributions corresponding to the mixture and BMA models of Sections \ref{subs3.1} and \ref{subs3.2}, respectively. In such cases, based on the representation on the right hand side of \eqref{eq:CRPSdef}, which also implies that the CRPS can be reported in the same units as the observation, one can consider the Monte Carlo approximation of the CRPS based on a large sample drawn from \ $F$ \ \citep[see e.g.][]{kltg21}. In the case studies of Section \ref{sec4}, the predictive performance of the various forecasts with a given lead time are compared using the mean CRPS over all forecast cases in the validation period.

Furthermore, the forecast skill of the competing forecasts with respect to  dichotomous events can be quantified with the help of the mean Brier score \citep[BS;][Section 9.4.2]{w19}. For a predictive CDF \ $F$ \ and the event that the observed visibility \ $x$ \ does not exceed a given threshold \ $y$, \ the BS is defined as
\begin{equation*}
\bs(F,x;y) := \big[F(y)-{\mathbb I}_{\{y\geq x\}}\big]^2,
\end{equation*}
so the CRPS is just the integral of the BS over all possible thresholds. 

For a probabilistic forecast \ $F$, \ one can assess the improvement with respect to a reference forecast \ $F_{\mathrm{ref}}$ \ in terms of a score  \ $\mathcal S$ \ by using the corresponding skill score \citep{murphy73}, defined as
\begin{equation*}
  \mathcal{SS}_F:=1-\frac {\overline{\mathcal S}_F}{\overline{\mathcal S}_{F_{\mathrm{ref}}}},
\end{equation*}
where \ $\overline{\mathcal S}_F$ \ and \ $\overline{\mathcal S}_{F_{\mathrm{ref}}}$ \ denote the mean score values corresponding  to forecasts 
 \ $F$ \ and  \ $F_{\mathrm{ref}}$, \ respectively. Skill scores are positively oriented (the larger the better), and in our case studies we report the continuous ranked probability skill score (CRPSS) and the Brier skill score (BSS).

Calibration and sharpness can also be investigated by examining the coverage and average width of \ $(1-\alpha )100\,\%, \ \alpha\in ]0,1[,$ \ central prediction intervals (intervals between the lower and upper \ $\alpha/2$ \ quantiles of the predictive distribution). Coverage is defined as the proportion of validating observations located in this interval and for a properly calibrated predictive distribution this value should be around \ $(1-\alpha )100\,\%$. \ Note that level \ $\alpha$ \ is usually chosen to match the nominal coverage of \ $(K-1)/(K+1)100\,\%$ \  of a $K$-member ensemble, which allows a direct comparison with the raw forecasts.

Further simple tools for assessing calibration of probabilistic forecasts are the verification rank histogram (or Talagrand diagram) of ensemble predictions and the probability integral transform (PIT) histogram of forecasts given in the form of predictive distributions. The Talagrand diagram is the histogram of ranks of the verifying observations with respect to the corresponding ensemble forecasts \citep[see e.g.][Section 9.7.1]{w19}, and in the case of a properly calibrated $K$-member ensemble the verification ranks should be uniformly distributed on \ $\{1,2,\ldots ,K+1\}$. \ The PIT is the value of the predictive CDF evaluated at the verifying observation with a possible randomization in the points of discontinuity \citep[see e.g.][Section 9.5.4]{w19}. The PIT values of calibrated predictive distributions follow a standard uniform law and in this way the PIT histogram can be considered as the continuous counterpart of the verification rank histogram.

Furthermore, the mean and the median of the preditive distributions and the ensemble mean and median as well can be considered as point forecasts for the corresponding weather variable. As the former optimizes the root mean squared error (RMSE), whereas the latter the  mean absolute error (MAE), we use these two scores are applied to evaluate the accuracy of point predictions \citep{g11}.

Finally, some of the the skill scores are accompanied with 95\,\% confidence intervals based on 2000 block bootstrap samples obtained using the stationary bootstrap scheme with mean block length derived according to \citet{pr94}. In this way one can  get insight into the uncertainty in verification scores and significance of score differences.

\section{Case studies}
\label{sec4}

The predictive performance of the novel mixture model introduced in Section \ref{subs3.1} is tested on the two datasets of ECMWF visibility ensemble forecasts and corresponding observations described in Section \ref{sec2}. As reference we consider the BMA approach provided in Section \ref{subs3.2}, climatological forecasts (observations of a given training period are considered as a forecast ensemble) and the raw ECMWF ensemble as well. Both parametric post-processing models require rather large training data to ensure reliable parameter estimation; moreover, seasonal variations of visibility should also be taken into account during the modelling process. In the case of the mixture model this latter requirement is addressed with the use of the annual base functions \eqref{eq:abf} in the locations of the component distributions. Hence, one can consider long training periods, which besides regional modelling allows clustering-based semi-local or even local parameter estimation, too. In the following sections regional, clustering-based semi-local and local mixture models are referred to as {\em Mixed-R\/}, {\em Mixed-C\/} and {\em Mixed-L\/}, respectively. In contrast to the mixture model, there is no seasonality included in the BMA predictive distribution, so short training periods are preferred allowing only regional modelling. BMA models of both Sections \ref{subs4.1} and \ref{subs4.2} are based on 25-day rolling training periods, which length is a result of a detailed data analysis (comparison of various BMA verification scores for a whole calendar year for training periods of 20, 25, 30, 35 and 40 days). Note that this training period length is identical to the one suggested by \citet{cr11}. Furthermore, as mentioned, for both calibration approaches separate modelling is performed for each lead time. Finally, in both case studies the size of the climatological forecasts matches the size of the corresponding raw ensemble predictions, that is in Section \ref{subs4.1}  observations of 51-day rolling training periods (see Section \ref{subs3.3}) are considered, whereas in Section \ref{subs4.2} climatology is based on 52 past observations. 

\begin{figure}[t]
\begin{center}
\epsfig{file=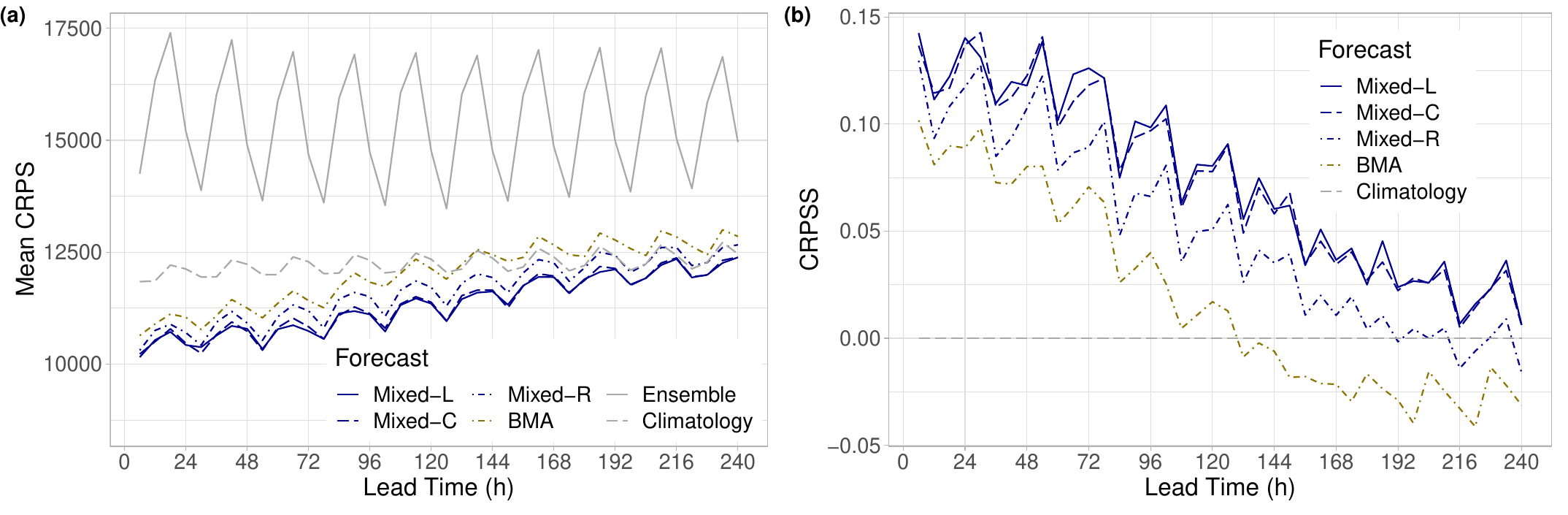, width=\textwidth}
\end{center}
\caption{Mean CRPS of post-processed, raw and climatological visibility forecasts for calendar year 2021 (a) and CRPSS of post-processed forecasts with respect to climatology (b) as functions of the lead time.}
\label{fig:crps_crpss}
\end{figure}

\begin{table}[t]
  \begin{center}
    \begin{tabular}{c|c|c|c|c|c|c}
  Mixed-L&Mixed-C&Mixed-R&BMA&Climatology\\ \hline
  73.44\,\%&73.60\,\%&75.41\,\%&77.85\,\%&79.34\,\%
 \end{tabular}
\end{center}
\caption{Overall mean CRPS of post-processed and climatological visibility forecasts for calendar year 2021 as proportion of the mean CRPS of the raw ECMWF ensemble.}
\label{tab1}
\end{table}

\subsection{Model performance for 51-member visibility ensemble forecasts}
\label{subs4.1}
In this case study the predictive performance of the competing forecasts is compared using data of calendar year 2021. For the 51-member ECMWF ensemble (control forecast and 50 exchangeable members) the mixture model has 15 free parameters to be estimated, and the comparison of the forecast skill of regional models based on training periods of length 100, 150, \ldots , 350 days reveals that the longest considered training period results in the best predictive performance. This 350-day training window is kept also for local and semi-local modelling, where the 13 locations are grouped into 6 clusters. Semi-local models with 3, 4 and 5 clusters have also been tested; however, these models slightly underperform the chosen one. Furthermore, as mentioned, the 11 parameters of the BMA model are estimated regionally using 25-day rolling training periods, which means a total of 325 forecast cases for each training step. Hence, the data/parameter ratio of the regional BMA approach (325/11=29.5) is slightly above of the corresponding ratio of the local mixture model (350/15=23.3). Note that in this case study validating visibility observations are reported up to 75 km, hence the support of all investigated post-processing models is limited to the 0 -- 75 km interval with a point mass at the upper bound.

\begin{figure}[t]
\begin{center}
\epsfig{file=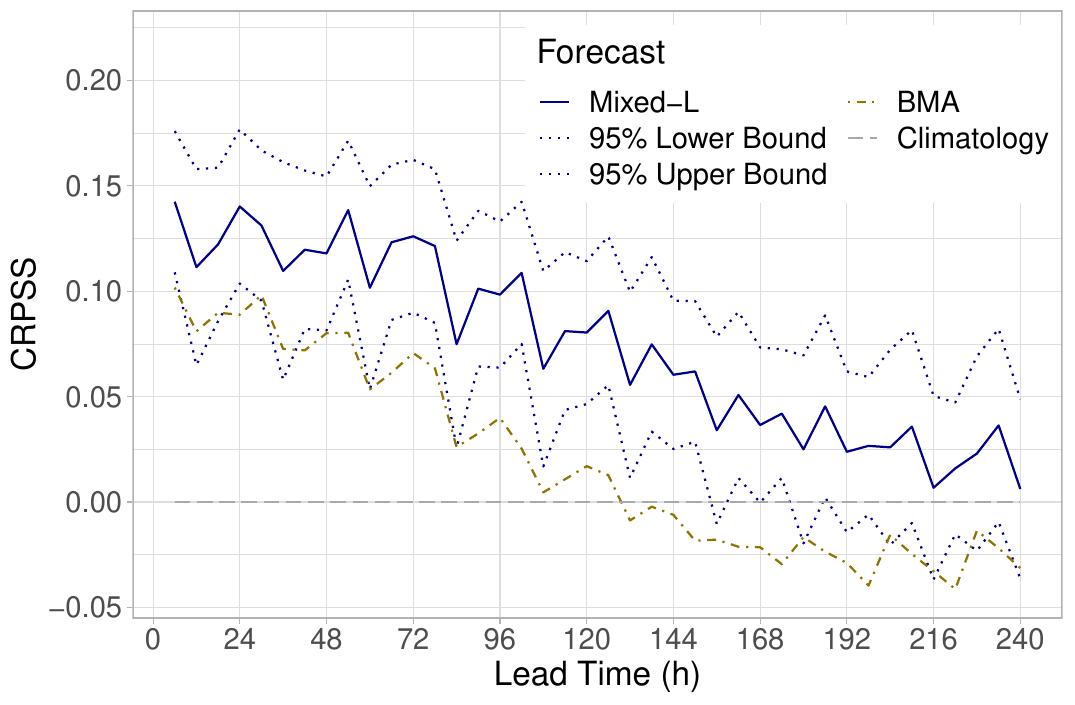, width=.5\textwidth}
\end{center}
\caption{CRPSS with respect to climatology of the best performing mixed model  (together with 95\,\% confidence intervals) and the BMA approach for calendar year 2021 as functions of the lead time.}
\label{fig:boot_crpss}
\end{figure}

Figure \ref{fig:crps_crpss}a indicates that in terms of the mean CRPS all investigated forecasts considerably outperform the raw visibility ensemble. Note, that the clearly recognizable oscillation in the CRPS can be explained by the four different observation times per day, and the raw ensemble exhibits the strongest dependence on the time of the day having the highest skill at 0600 UTC. Parametric models are superior to climatology only for shorter forecast horizons and their advantage gradually fades with the increase of the lead time. The difference between post-processed forecasts and climatology is more visible in the CRPSS values of Figure \ref{fig:crps_crpss}b. Skill scores of the locally and semi-locally trained mixture models are positive for all lead times and the difference between these forecasts is negligible. Up to 192 h the Mixed-R approach also outperforms climatology and it is clearly ahead of the BMA, which results in positive CRPSS only for shorter forecast horizons (6 -- 126 h). This ranking of the competing methods is also confirmed by Table \ref{tab1} providing the overall mean CRPS values of calibrated and climatological forecasts as proportions of the mean CRPS of the raw ECMWF visibility ensemble. We remark, that a similar good performance of climatology with respect to raw and BMA post-processed visibility forecasts of the University of Washington Mesoscale Ensemble was observed by \citet{cr11}. 

\begin{figure}[t]
\begin{center}
\epsfig{file=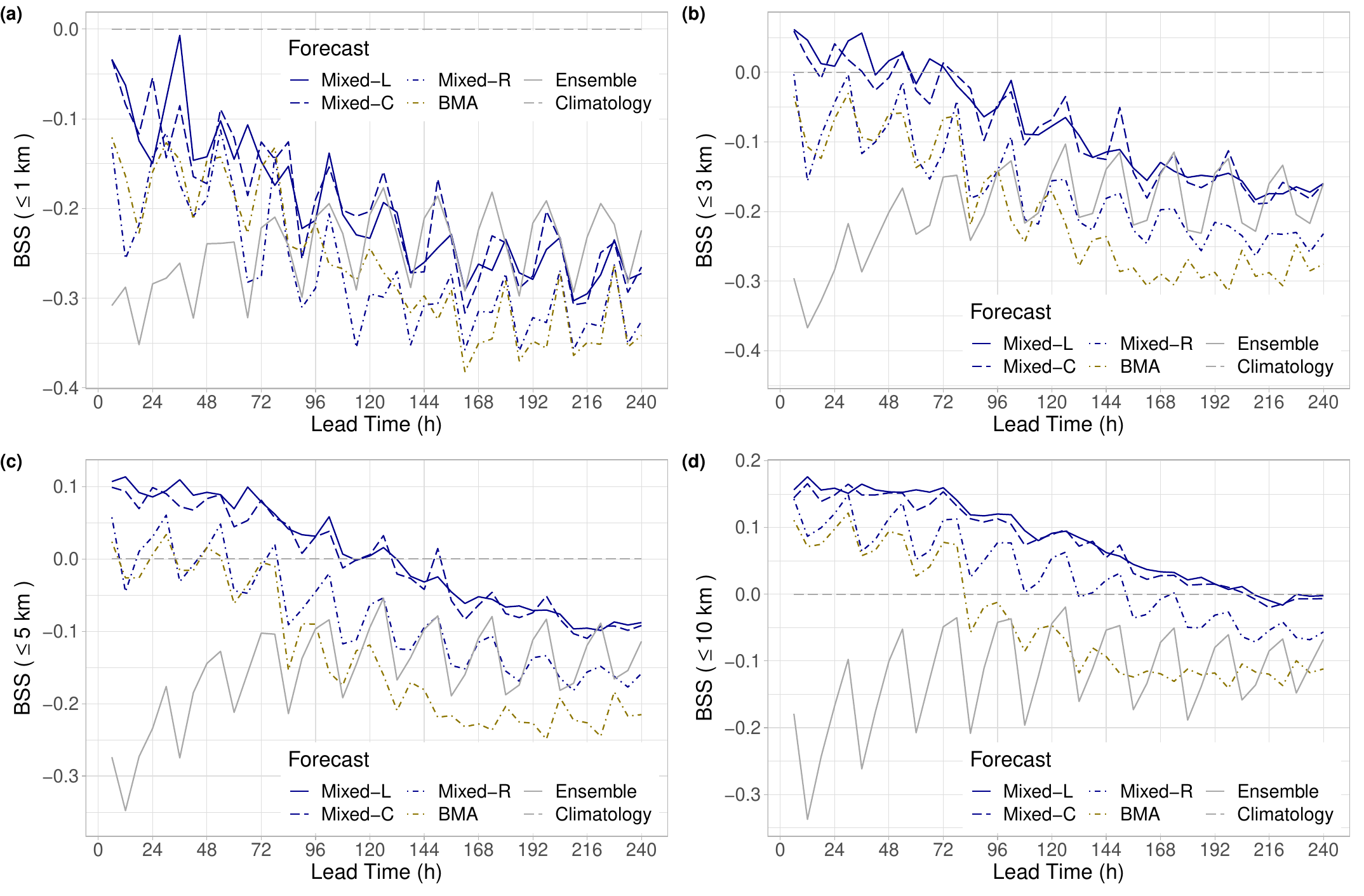, width=\textwidth}
\end{center}
\caption{BSS of raw and post-processed visibility forecasts for calendar year 2021 with respect to climatology for thresholds 1 km (a), 3 km (b), 5 km (c) and 10 km (d) as functions of the lead time.}
\label{fig:bss}
\end{figure}

In Figure \ref{fig:boot_crpss} the CRPSS values of the best performing locally trained mixture model are accompanied with 95\,\% confidence intervals, which helps in assessing the significance of the differences in CRPS. The superiority of the Mixed-L forecast over climatology in terms of the mean CRPS is significant at a 5\,\% level up to 150 h and between 42 h and 174 h significantly outperforms the BMA model as well. We remark that for the latter approach after 96 h the CRPSS with respect to climatology fails to be significantly positive (not shown).

\begin{figure}[h!]
\begin{center}
  \epsfig{file=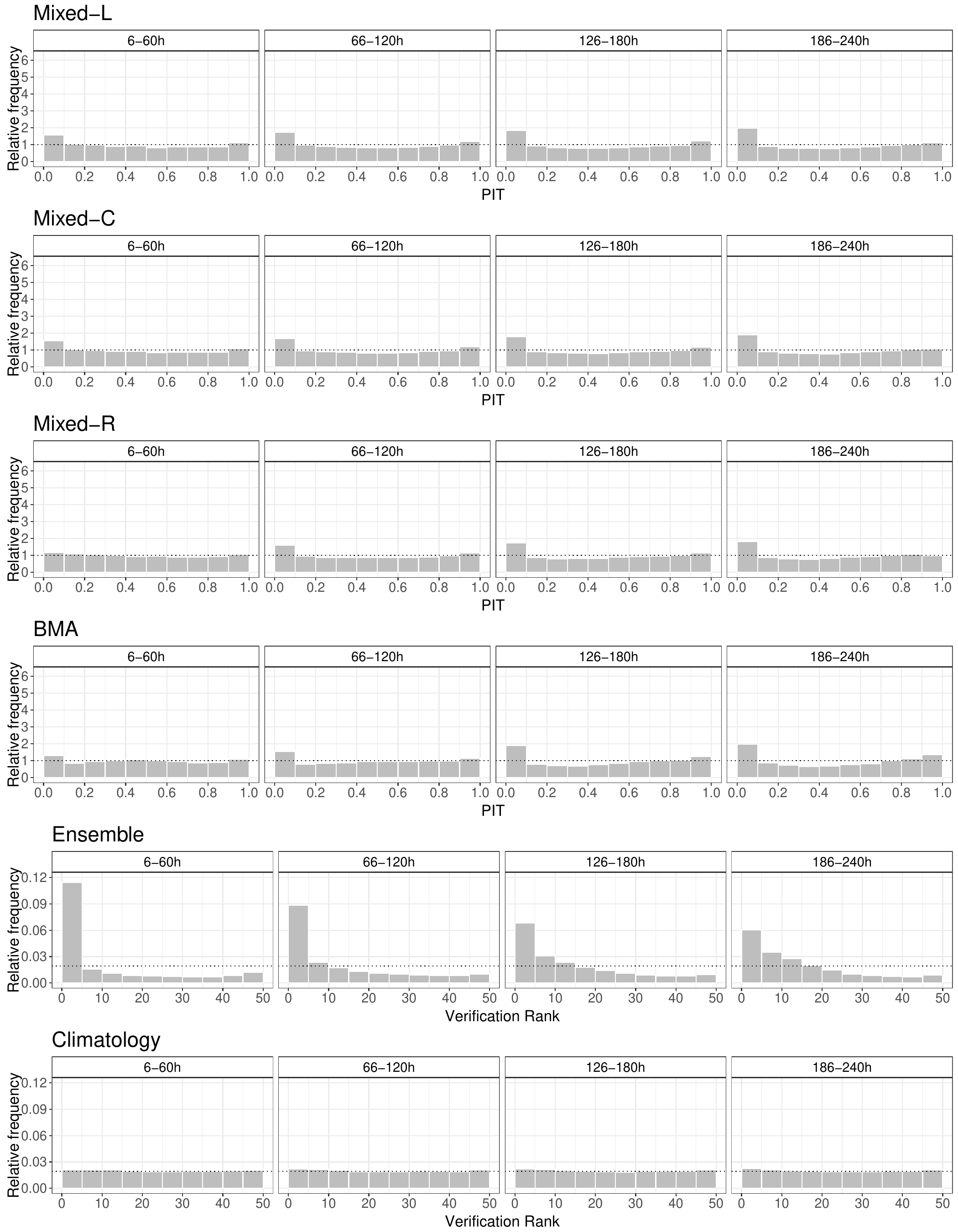, width=.95\textwidth}
\end{center}
\caption{ PIT histograms of post-processed and verification rank histograms of climatological and raw visibility forecasts for calendar year 2021 for lead times 6–60 h, 66–120 h, 126–180 h and 186–240 h.}
\label{fig:pit}
\end{figure}

The analysis of Brier skill scores plotted in Figure \ref{fig:bss} slightly tones the picture about the performance of post-processed forecasts. For visibility not exceeding 1 km, none of the competitors outperforms climatology (Figure \ref{fig:bss}a) and calibrated predictions are superior to raw ensemble forecasts only for short lead times. With the increase of the threshold the positive effect of post-processing is getting more and more pronounced and the ranking of the different models starts matching the one based on the mean CRPS. From the competing calibration methods the locally and semi-locally trained mixed models consistently display the highest skill and for the largest threshold value of 10 km they outperform climatology up to 204 h (Figure \ref{fig:bss}d).

\begin{figure}[t]
\begin{center}
\epsfig{file=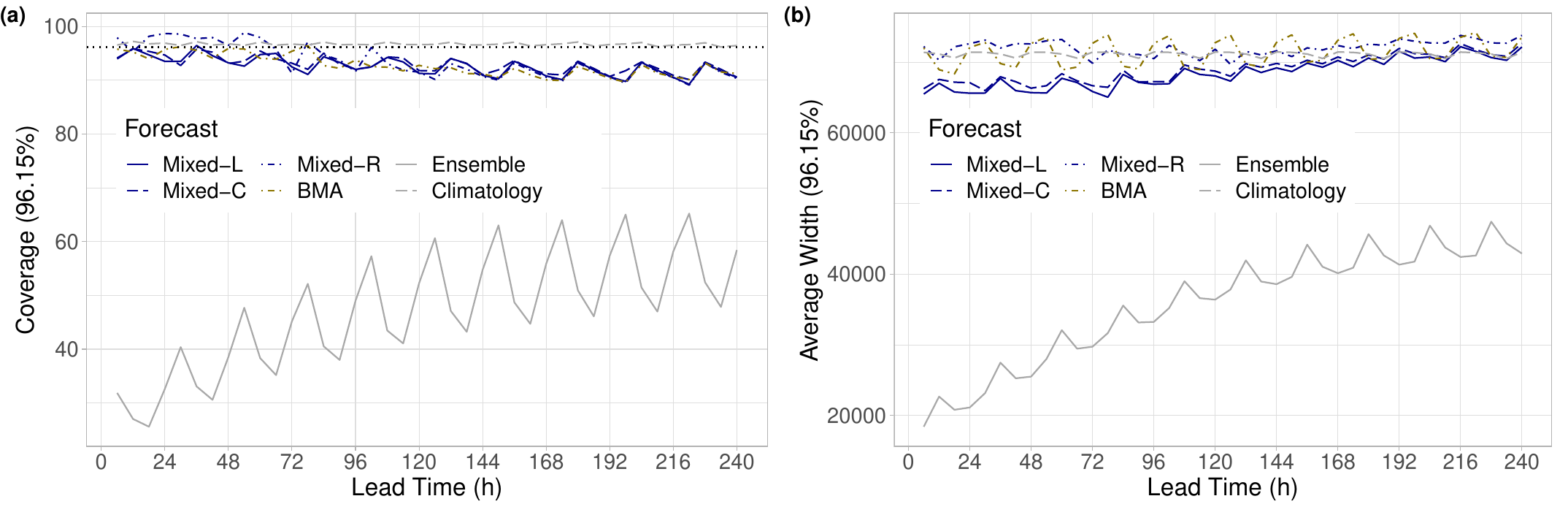, width=\textwidth}
\end{center}
\caption{Coverage (a) and average width (b) of nominal 96.15\,\% central prediction intervals of raw and post-processed visibility forecasts for calendar year 2021 as functions of the lead time. In panel (a) the ideal coverage is indicated by the horizontal dotted line.}
\label{fig:cov_aw}
\end{figure}

The verification rank and PIT histograms of Figure \ref{fig:pit} again illustrate the primacy of climatology over the raw ensemble and the improved calibration of post-processed forecasts. Raw ECMWF visibility forecasts are underdispersive and tend to overestimate the observed visibility; however, these defficiencies improve with the increase of the forecast horizon. Climatology results in almost uniform rank histograms with just a minor underdispersion and there is no visible dependence on the forecast horizon. Unfortunately, none of the four investigated post-processing models can completely eliminate the bias of the raw visibility forecasts, which is slightly more pronounced for longer lead times.

\begin{figure}[t]
\begin{center}
\epsfig{file=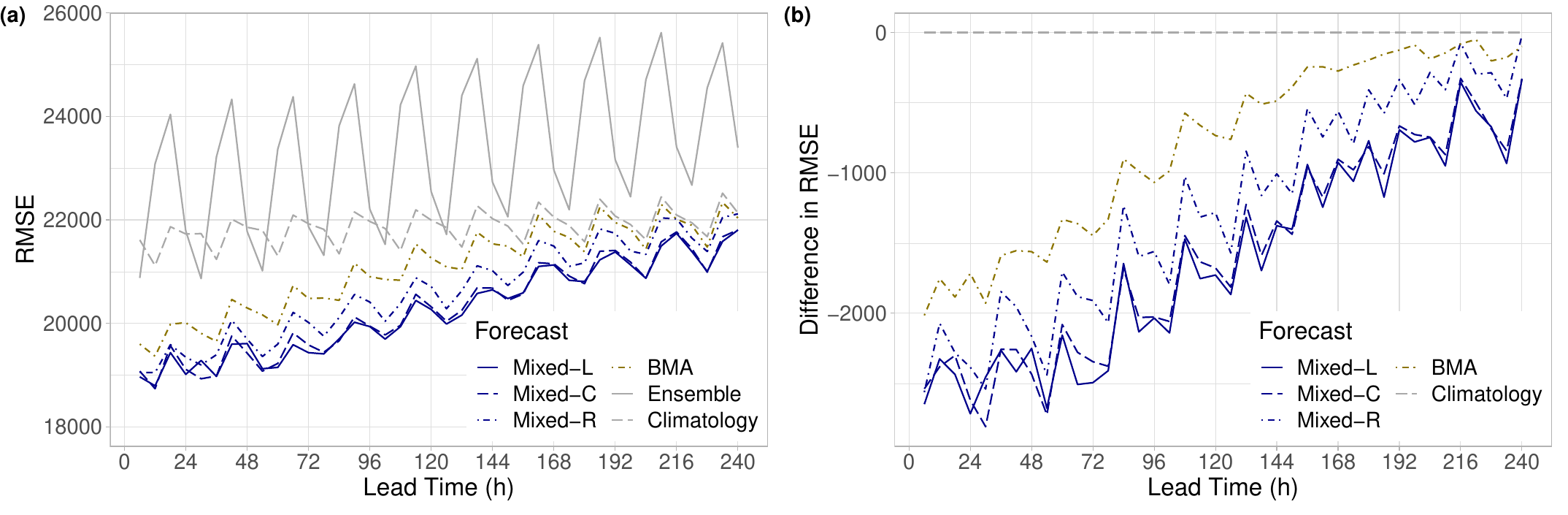, width=\textwidth}
\end{center}
\caption{RMSE of the mean forecasts for calendar year 2021 (a) and difference in RMSE from climatology (b) as functions of the lead time.}
\label{fig:rmse_rmsed}
\end{figure}

Furthermore, the coverage values of nominal 96.15\,\% central prediction intervals depicted in Figure \ref{fig:cov_aw}a are fairly consistent with the shapes of the corresponding verification rank and PIT histograms. The underdispersion of the raw ensemble is confirmed with its low coverage, which shows an increasing trend, a clear diurnal cycle, and ranges from 25.58\,\% to 65.21\,\%. As one can observe on the corresponding curve of Figure \ref{fig:cov_aw}b, the improvement of the ensemble coverage with the increase of the forecast horizon is a consequence of the increase in spread, that results in expanding central prediction intervals. The price of the almost perfect coverage of climatology with a mean absolute deviation from the nominal value of 0.53\,\% is the much wider central prediction interval. The coverage values of post-processed forecasts decrease with the increase of the lead time, the corresponding mean absolute deviations from the nominal 96.15\,\% are 3.55\,\% (Mixed-L), 3.21\,\% (Mixed-C), 3.55\,\% (Mixed-R) and 3.39\,\% (BMA). However, this negative trend in coverage, especially in the case of the locally and semi-locally trained mixed models, is combined with increasing average width, which can be a consequence of the growing bias.

Finally, in terms of the RMSE of the mean forecast, all post-processing approaches outperform both the raw ensemble and climatology for all lead times (see Figure \ref{fig:rmse_rmsed}a); however, their advantage over climatology is negatively correlated with the forecast horizon. Mixed-L and Mixed-C approaches result in the lowest RMSE values, followed by the Mixed-R and BMA forecasts (see also Figure \ref{fig:rmse_rmsed}b), which order perfectly matches the ranking based on the mean CRPS (Figure \ref{fig:crps_crpss}b) and the mean BS for all studied thresholds (Figure \ref{fig:bss}).

\begin{figure}[t]
\begin{center}
\epsfig{file=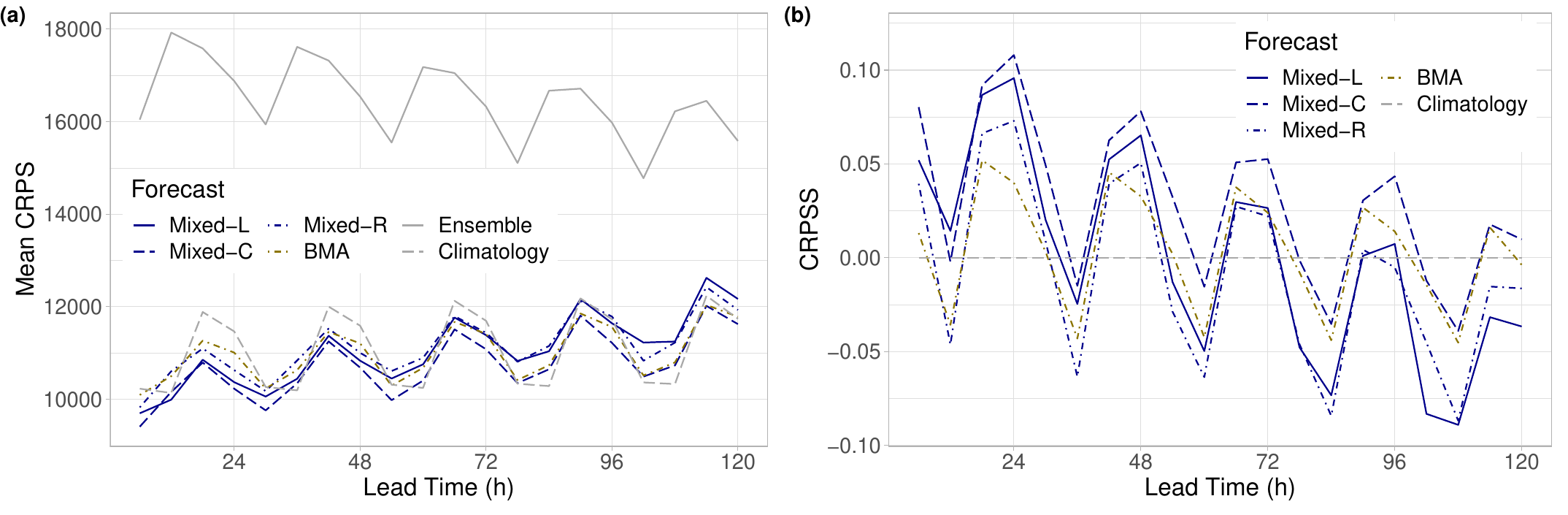, width=\textwidth}
\end{center}
\caption{Mean CRPS of post-processed, raw and climatological EUPPBench visibility forecasts for calendar year 2018 (a) and CRPSS of post-processed forecasts with respect to climatology (b) as functions of the lead time.}
\label{fig:crps_crpssB}
\end{figure}

\begin{table}[t]
  \begin{center}
    \begin{tabular}{c|c|c|c|c|c|c}
  Mixed-L&Mixed-C&Mixed-R&BMA&Climatology\\ \hline
  66.90\,\%&64.96\,\%&67.41\,\%&66.80\,\%&67.28\,\%
 \end{tabular}
\end{center}
\caption{Overall mean CRPS of post-processed and climatological EUPPBench visibility forecasts for calendar year 2018 as proportion of the mean CRPS of the raw ECMWF ensemble.}
\label{tab2}
\end{table}

\subsection{Model performance for EUPPBench visibility ensemble forecasts}
\label{subs4.2}

Since in the EUPPBench benchmark dataset the 51-member ECMWF ensemble forecast is augmented with the deterministic high-resolution prediction, the mixture model \eqref{eq:mixPDF} has 17 free parameters to be estimated, whereas for the BMA predictive PDF \eqref{eq:bma}, the parameter vector is 16 dimensional. As mentioned, BMA modelling is based on 25-day regional training, while in order to determine the optimal training period length for mixed models, we again compare the skill of the regionally estimated forecasts based on rolling training windows of 100, 150, \ldots, 350 days. In the case of the EUPPBench visibility data the skill of the different models is compared with the help of forecast-observation pairs for calendar year 2018. In contrast to the previous case study, where the longest tested training period of 350 days is preferred, here the 100-day window results in the best overall performance. Using the same training period length we again investigate local modelling and semi-local estimation based on 4 clusters. Taking into account the maximal reported visibility observation in the EUPPBench benchmark dataset, now the mixed and BMA predictive distributions have point masses at 70 km.

Again, Figure \ref{fig:crps_crpssB}a shows the mean CRPS of post-processed, raw and climatological EUPPBench visibility forecasts  as functions of the lead time, while in Figure \ref{fig:crps_crpssB}b the CRPSS values of the mixed and BMA models with respect to the 52-day climatology are plotted. Similar to the case study of Section \ref{subs4.1}, climatological and post-processed forecasts outperform the raw ensemble by a wide margin; however, now the advantage of post-processing over climatology is not so obvious and the ranking of the calibration methods also differ. From the four investigated models the Mixed-C model results in the lowest mean CRPS for all lead times but 12 h; nevertheless, even this approach shows negative skill against climatology for lead times corresponding to 1200 UTC observations. Local modelling (Mixed-L) is competitive only for short forecast horizons, which might be explained by the short training period leading to numerical issues during parameter estimation due to the low data/parameter ratio. In general, the skill scores of all post-processing methods show a decreasing trend with the BMA having the mildest slope. Based on Table \ref{tab2}, providing the improvement in the overall mean CRPS over the raw ECMWF ensemble, one can establish ranking Mixed-C -- BMA -- Mixed-L -- Climatology -- Mixed-R. Note that the improvements provided here are much larger then the ones in Table \ref{tab1} and in terms of the mean CRPS the 52-member EUPPBench visibility forecasts are behind the more recent 51-member ensemble predictions studied in Section \ref{subs4.1}. This dissimilarity in forecast performance is most likely due to the consecutive improvement in the ECMWF IFS; however, it might also be related to the  difference in the forecast domains (see Figure \ref{fig:map}).

\begin{figure}[t]
\begin{center}
\epsfig{file=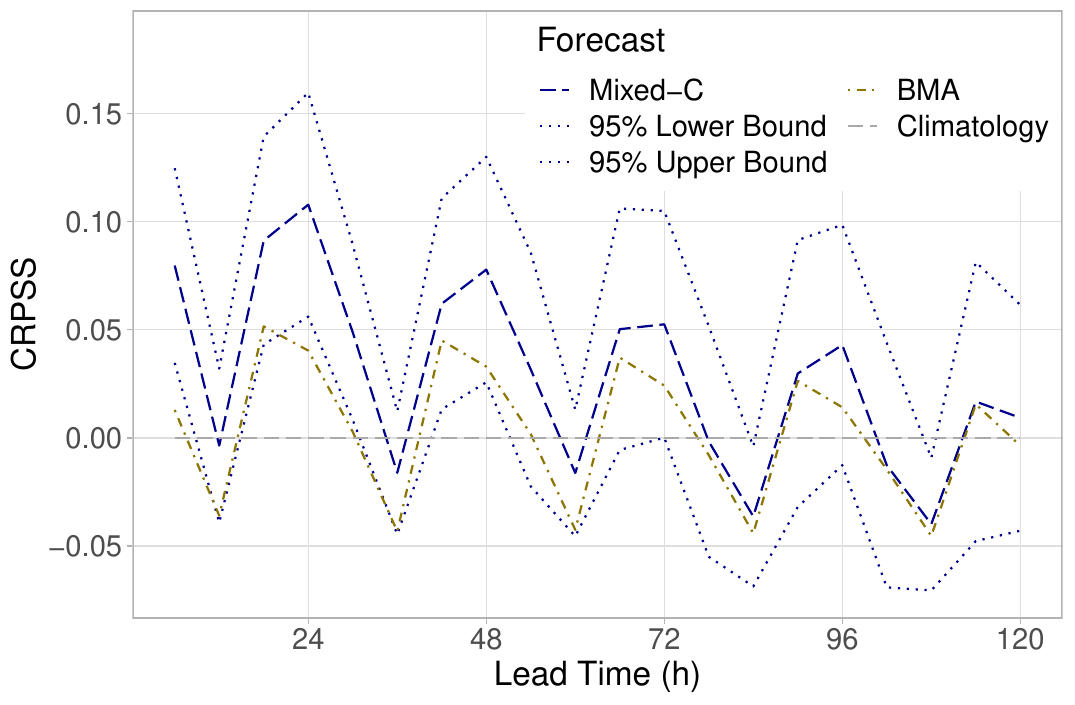, width=.5\textwidth}
\end{center}
\caption{CRPSS with respect to climatology of the best performing mixed model  (together with 95\,\% confidence intervals) and the BMA approach for calendar year 2018 as functions of the lead time.}
\label{fig:boot_crpssB}
\end{figure}

Furthermore, according to Figure \ref{fig:boot_crpssB}, even for 0000, 0600 and 1800 UTC observations the advantage of the best performing Mixed-C model over climatology is significant at a 5\,\% level only up to 48 h, whereas the difference in skill from the BMA approach is significant just at 6 h, 24 h and 30 h. Note that the CRPSS of the BMA model with respect to climatology is significantly positive at a 5\,\% only for lead times 18 h, 24 h, 40 h and 48 h (not shown).

\begin{figure}[t]
\begin{center}
\epsfig{file=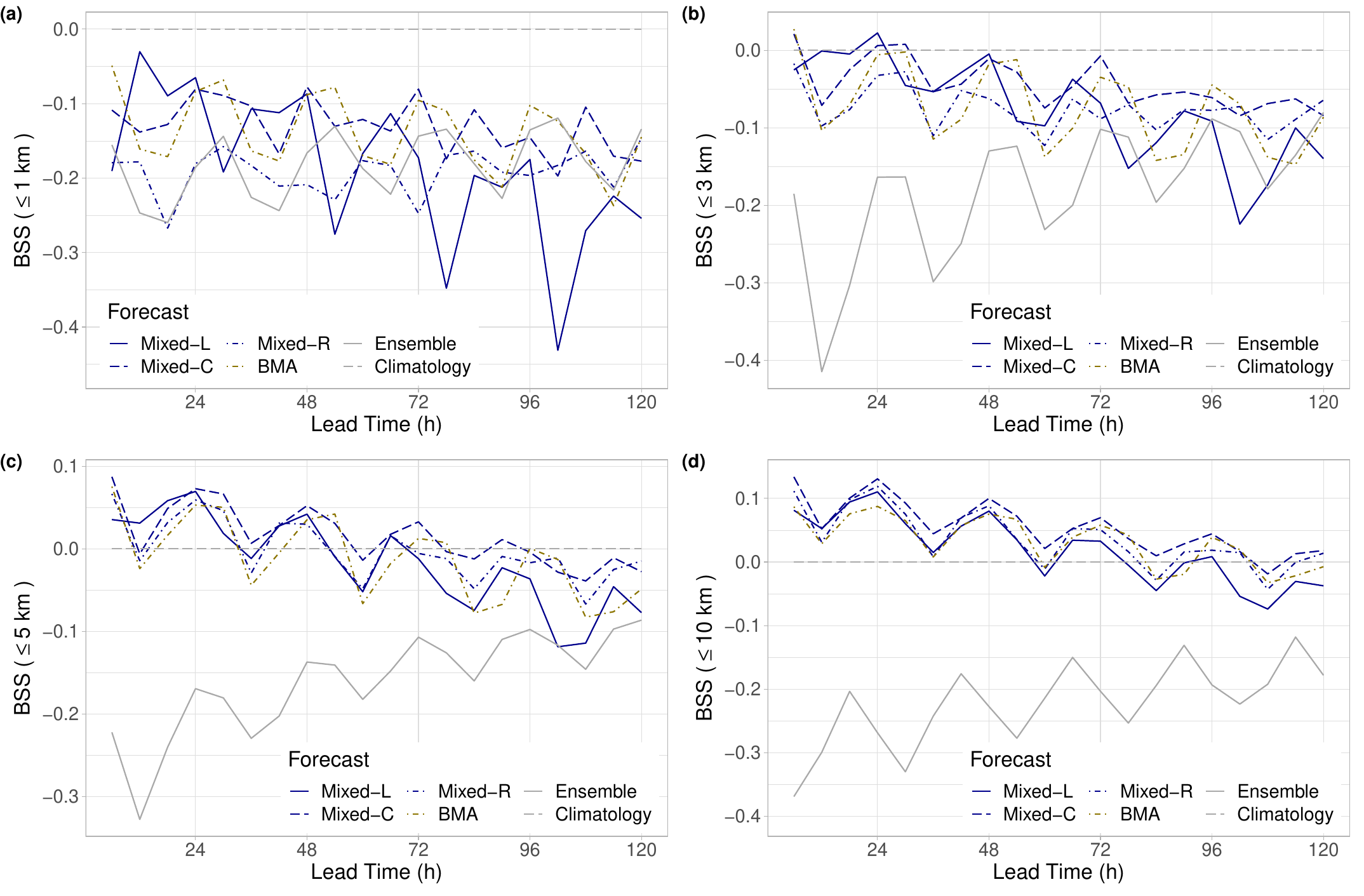, width=\textwidth}
\end{center}
\caption{BSS of raw and post-processed EUPPBench visibility forecasts for calendar year 2018 with respect to climatology for thresholds 1 km (a), 3 km (b), 5 km (c) and 10 km (d) as functions of the lead time.}
\label{fig:bssB}
\end{figure}

The Brier skill scores of Figure \ref{fig:bssB} lead us to similar consequences as in the previous case study. For the lowest threshold of 1 km, all forecasts underperform climatology (Figure \ref{fig:bssB}a); however, the skill of post-processed predictions improves when the threshold is increased. For 3, 5 and 10 km threshold the ranking of the various models is again identical to the ordering based on the mean CRPS (see Figure \ref{fig:crps_crpssB}b) with the Mixed-C approach exhibiting the best overall predictive performance, closely followed by the BMA model. For the largest threshold of 10 km, up to 54 h, climatology is outperformed even by the least skillful Mixed-L approach (see  Figure \ref{fig:bssB}d), whereas the leading semi-locally trained mixed model results in a positive BSS up to 102 h.

\begin{figure}[h!]
\begin{center}
  \epsfig{file=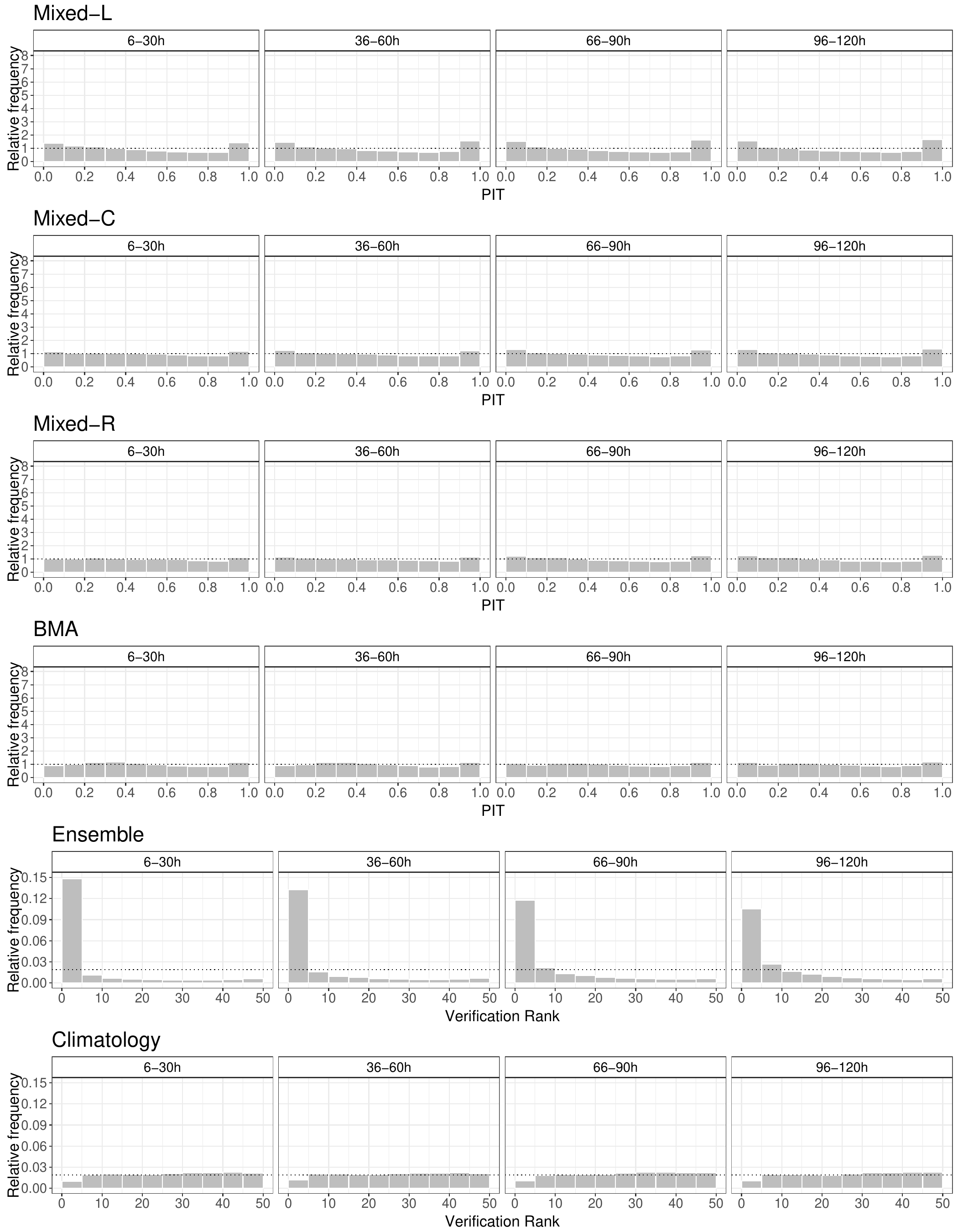, width=0.95\textwidth}
\end{center}
\caption{ PIT histograms of post-processed and verification rank histograms of climatological and raw EUPPBench visibility forecasts for calendar year 2018 for lead times 6–30 h, 36–60 h, 66–90 h and 96–120 h.}
\label{fig:pitB}
\end{figure}

The verification rank histograms of the raw EUPPBench visibility forecasts depicted in Figure \ref{fig:pitB} show a much stronger bias than the corresponding panel of Figure \ref{fig:pit}, while the improvement with the increase of the forecast horizon is less pronounced. Climatology is also slightly biased but in the opposite direction, whereas the verification rank histograms of all post-processed forecasts are closer to the desired uniform distribution than in the case study of Section \ref{subs4.1}. Here the locally trained mixed model exhibits the strongest bias; however, neither the verification rank histograms of climatology, nor the PIT histograms of the calibrated forecasts indicate visible dependence on the forecast horizon. 
\begin{figure}[t]
\begin{center}
\epsfig{file=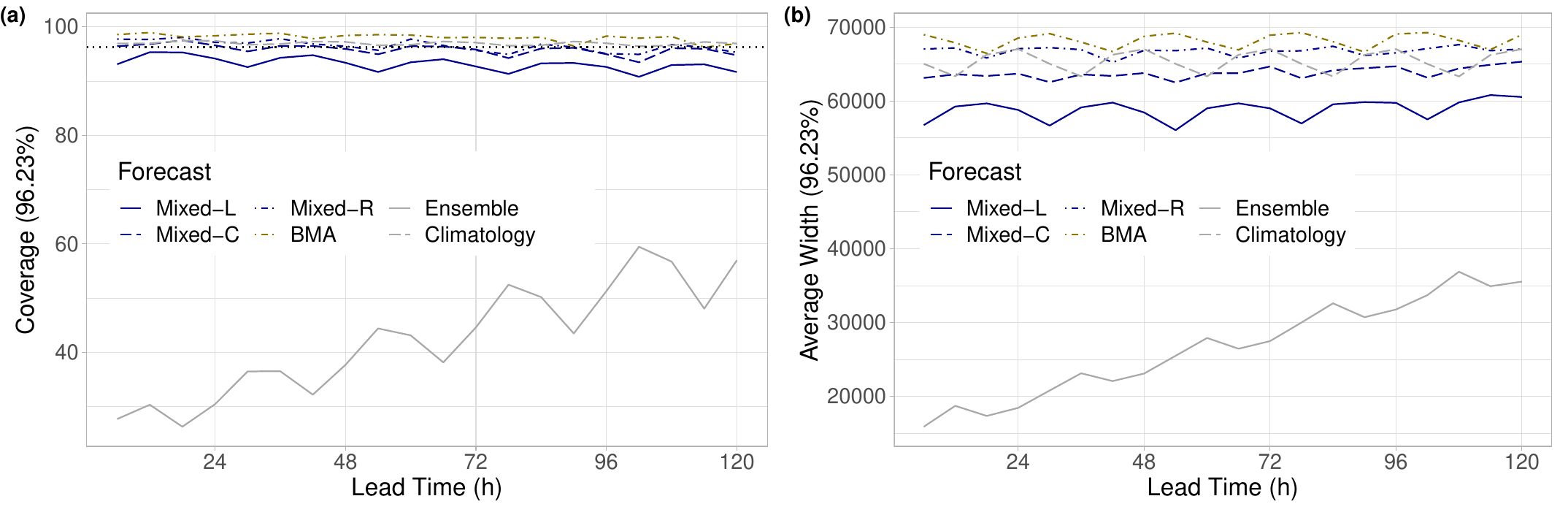, width=\textwidth}
\end{center}
\caption{Coverage (a) and average widths (b) of nominal 96.23\,\% central prediction intervals of raw and post-processed EUPPBench visibility forecasts for calendar year 2018 as functions of the lead time. In panel (a) the ideal coverage is indicated by the horizontal dotted line.}
\label{fig:cov_awB}
\end{figure}

The fair calibration of climatological and post-processed forecasts can also be observed in Figure \ref{fig:cov_awB}a displaying the coverage values of the nominal 96.13\,\% central prediction intervals. Semi-locally and regionally trained mixed models and climatology result in almost perfect coverage, closely followed by the BMA model; the corresponding mean absolute deviations from the nominal value are 0.69\,\%, 0.90\,\%, 0.72\,\% and 1.81\,\%, respectively. The Mixed-L model is slightly behind its competitors with a mean absolute deviation of 3.00\,\%, whereas the maximal coverage of the raw EUPPBench ensemble does not reach 60\,\%. Note that the ranking of the various predictions, the increasing coverage of the raw ensemble, and the lack of visible trend in the coverage values of post-processed forecasts and climatology is pretty much in line with the shapes of the corresponding histograms of Figure \ref{fig:pitB}.
In general, the average widths of the investigated 96.13\,\% central prediction intervals (Figure \ref{fig:cov_awB}b) are rather consistent with the matching coverage values. Nevertheless, one should remark that the best performing Mixed-C model results in sharper predictions than climatology and the Mixed-R and BMA approaches.
 
\begin{figure}[t]
\begin{center}
\epsfig{file=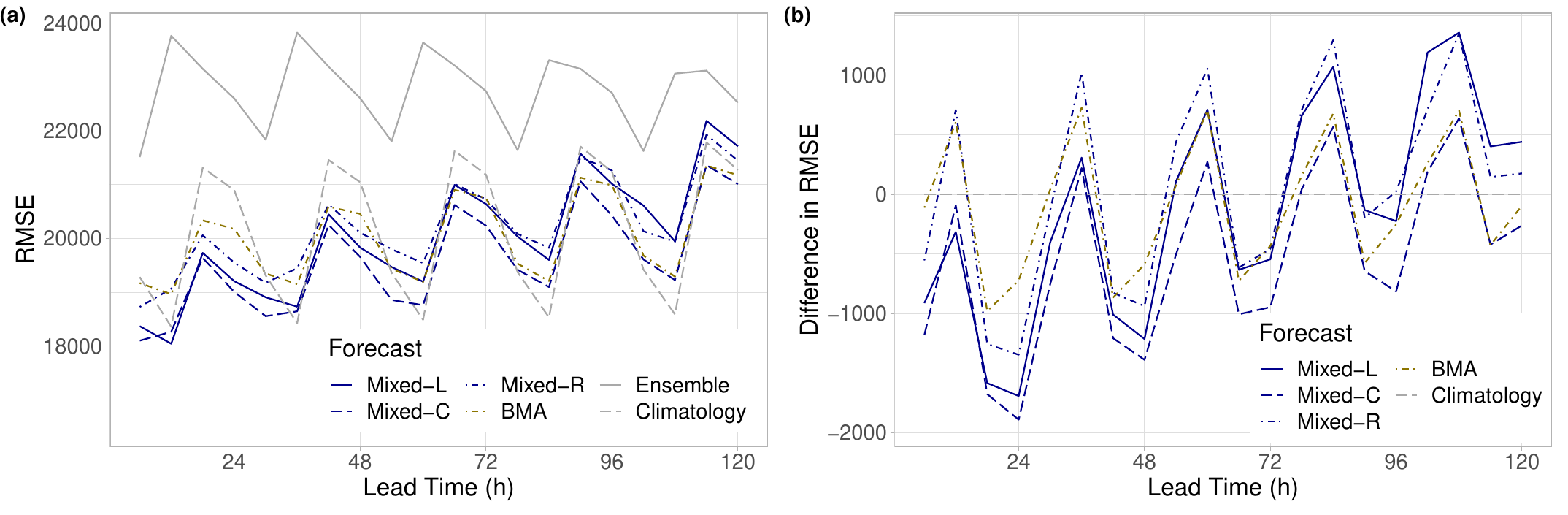, width=\textwidth}
\end{center}
\caption{RMSE of the mean EUPPBench forecasts for calendar year 2018 (a) and difference in RMSE from climatology (b) as functions of the lead time.}
\label{fig:rmse_rmsedB}
\end{figure}

Finally, according to Figure \ref{fig:rmse_rmsedB}, in terms of the RMSE of the mean, we see a similar behaviour and ranking of the different forecasts as in the case of the mean CRPS (see Figure \ref{fig:crps_crpssB}). While the raw ensemble is clearly behind the other forecasts, with the increse of the lead time climatology becomes more and more competitive, especially at forecast horizons corresponding to 1200 UTC observations. However, up to 30 h, both locally and semi-locally trained mixed models result in lower RMSE than the climatological forecast, and the Mixed-C approach consistently outperforms all other calibration methods for all lead times but 12 h.

\section{Conclusions}
\label{sec5}

We propose a novel parametric approach to calibrating visibility ensemble forecasts, where the predictive distribution is a mixture of a gamma and a truncated normal law, both right censored at the maximal reported visibility. Three model variants that differ in the spatial selection of training data are evaluated in two case studies, where as reference post-processing method we consider the BMA model of \citet{cr11}; however, we also investigate the skill of climatological and raw ensemble forecasts. While both case studies are based on ECMWF visibility predictions with a 6 h temporal resolution, they cover distinct geographical regions and time intervals, and only one of them uses the deterministic high resolution forecast. The results presented in Section \ref{sec4} indicate, that all post-processing models consistently outperform the raw ensemble by a wide margin and the real question is whether statistical calibration results in improvement compared to climatology. In the case of the 51-member operational ECMWF ensemble, e.g. in terms of the mean CRPS of the probabilistic and RMSE of the mean forecasts, the best performing locally and semi-locally trained mixed models outperform climatological predictions for all investigated lead times. For the EUPPBench dataset the situation is far from being so obvious; post-processing can result consistently positive skill with respect to climatology only up to 30 h. In general, the advantage of post-processed forecasts over climatology shows a decreasing trend with the increase of the forecast horizon, locally and semi-locally trained mixed models are preferred against the regionally estimated one, and the BMA approach is slightly behind the competitors. Note that the general conclusions about the effect of post-processing and the behaviour and ranking of the raw, climatological and calibrated visibility forecasts are almost completely in line with the results of \citet{bl23}, where classification-based discrete post-processing of visibility is studied based on extended versions of the current visibility datasets (more observation stations from the same geographical regions).

The results of this study suggest several further directions of future research. One possible option is to consider a matching distributional regression network (DRN) model, where the link functions connecting the parameters of the mixture predictive distribution with the ensemble forecast are replaced by an appropriate neural network. This parametric machine learning-based approach is proved successful for several weather quantities, such as temperature \citep{rl18}, precipitation \citep{gzshf21}, wind gust \citep{sl22}, wind speed \citep{bb21} or solar irradiance \citep{bb24}.

Furthermore, one can also investigate the impact of introduction of additional covariates on the forecast skill of parametric models based on the proposed censored gamma -- truncated and censored normal mixture predictive distribution. In the DRN setup this step is rather straightforward and might result in significant improvement in predictive performance \citep[see e.g][]{rl18,sl22}. A natural choice can be any further visibility forecast (for instance, the one of the Copernicus Atmospheric Monitoring Service); however, forecasts of other weather quantities affecting visibility can also be considered.

Finally, using two-step multivariate post-processing techniques one can extend the proposed mixture model in order to obtain spatially and/or temporally consistent calibrated visibility forecasts. For an an overview of the state-of-the-art multivariate approaches we refer to \citet{lbm20} and \citet{llhb23}.

\bigskip
\noindent
{\bf Acknowledgments.} \ The work leading to this paper was done in part during the visit of S\'andor Baran to the Heidelberg Institute for Theoretical Studies in July 2023 as guest researcher. Both authors gratefully acknowledge the support of the the National Research, Development and Innovation Office under Grant No. K142849. Finally, the authors are indebted to Zied Ben Bouall\`egue for providing the ECMWF visibility data for 2020--2021.

\end{document}